\documentclass[conference]{IEEEtran}

\IEEEoverridecommandlockouts                              

\usepackage{epsfig}

\usepackage{breakurl}
\usepackage[breaklinks]{hyperref}
\usepackage{amsmath}
\usepackage{amsfonts}

\usepackage{amsthm}
\usepackage{amssymb}
\usepackage{graphicx}
\usepackage{color}
\usepackage{balance} 
\usepackage{bm}
\usepackage{siunitx}

\newcommand{\bma}{\begin{bmatrix}}
\newcommand{\ebma}{\end{bmatrix}}

\usepackage{enumitem}
\usepackage{graphicx} 
\graphicspath{{figure/}}
\usepackage{float}
\usepackage[caption=false]{subfig}
\usepackage{soul}
\usepackage{color}

\usepackage{cite}

\usepackage{algorithm,algpseudocode}

\newtheoremstyle{bfnote}%
{}{}%
{\itshape}{}%
{\bfseries}{.}%
{ }%
{\thmname{#1}\thmnumber{ #2}\thmnote{ (#3)}}
\theoremstyle{bfnote}

\usepackage{circuitikz}
\usepackage{tikz}

\title{
Safe Trajectory Gradient Flow Control of a Grid-Interfacing Inverter
\thanks{The authors were partially supported by NSF grant ECCS-2023531.}
}

\author{
	\IEEEauthorblockN{
		Trager Joswig-Jones, 
        Baosen Zhang
	}
	\IEEEauthorblockA{
		Department of Electrical and Computer Engineering, University of Washington, Seattle, WA, USA\\
        Emails: joswitra@uw.edu, zhangbao@uw.edu
	}
}

\begin{document}

\maketitle
\thispagestyle{empty}
\pagestyle{empty}

\begin{abstract}

Grid-interfacing inverters serve as the interface between renewable energy resources and the electric power grid, offering fast, programmable control capabilities. However, their operation is constrained by hardware limitations, such as bounds on the current magnitude. Existing control methods for these systems often neglect these constraints during controller design and instead rely on ad hoc limiters, which can introduce instability or degrade performance. In this work, we present a control framework that directly incorporates constraints into the control of a voltage-source inverter. We propose a safe trajectory gradient flow controller, which applies the safe gradient flow method to a rolling horizon trajectory optimization
problem to ensure that the states remain within a safe set defined by the constraints while directing the trajectory towards an optimal equilibrium point of a nonlinear program. Simulation results demonstrate that our approach can drive the outputs of a simulated inverter system to optimal values and maintain state constraints, even when using a limited number of optimization steps per control cycle. 

\end{abstract}
\vspace{-12pt}
%
%
\section{Introduction}
\label{sec:intro}


As renewable energy resources continue to gain prominence in electric power systems, power electronic inverters are increasingly replacing synchronous generators as the primary interface between energy sources and the grid~\cite{kroposki2017achieving}. Unlike traditional generators, inverters are fast, flexible, and programmable. However, this flexibility comes with strict physical constraints, particularly limits on current magnitude, that must be enforced to prevent damage to the devices. 


A number of control approaches have been proposed with the aim of supporting the stability of the grid, such as droop control~\cite{Chandorkar_Divan_Adapa_1993}, virtual synchronous machines~\cite{driesen2008virtual}, and virtual oscillator control~\cite{dhople2013virtual}. However, these methods often assume ideal actuation and neglect the physical limitations of inverter hardware. To compensate, auxiliary modules such as current limiters and virtual impedance methods are used, but these can introduce undesirable side effects, such as altered voltage dynamics or poor stability~\cite{Fan_Liu_Zhao_Wu_Wang_2022}. To address these limitations, optimization-based methods that directly consider the constraints of the inverter system have been introduced. A safety filter based approach is proposed in~\cite{joswig2024safe}, which minimally modifies a nominal control action to safely limit the current output, but does not consider the optimality of the control actions. 
The approach in~\cite{Groß_Lyu_2023} uses primal-dual gradient dynamics to solve a one-step optimal control problem under constraints, but does not guarantee constraint satisfaction at all times due to the use of quasi-steady state circuit models in its formulation.

Optimization-based control methods that steer a control system to a solution of an optimization problem have also been studied more broadly in the context of constrained dynamical systems~\cite{HAUSWIRTH2024100941}. Recent works have proposed online feedback controllers that continuously regulate a system to the solution of an optimization problem with constraints on the state at steady-state~\cite{Chen_Cothren_Cortés_Dall’Anese_2023, 8673636}. However, applying these methods to dynamic systems may lead to constraint violations during the transient. To solve this issue the approach in~\cite{Delimpaltadakis_Mestres_Cortés_Heemels_2025} uses high-order control barrier functions with safe gradient flows to ensure their feedback-optimization method enforces state and input constraints at all times. 

In this paper, we similarly build upon the safe gradient flow approach~\cite{cortes_anytime_2021} and develop a control strategy for discrete-time dynamical systems, specifically targeting the control of a grid-interfacing inverter. Our method casts the inverter control problem as a constrained trajectory optimization problem over a finite horizon and applies a rolling horizon approach similar to model predictive control. At each step, we perform a limited number of optimization updates using safe gradient flows, then apply only the first control action. This structure allows us to handle nonlinear dynamics, enforce safety constraints such as current limits, and operate without a full solution of the trajectory optimization problem at every step. Simulation results show that our method successfully regulates the inverter system to an optimal equilibrium while maintaining safety.


\section{Preliminaries}
\label{sec:prelim}
We introduce the notation used in the paper and recall definitions of control barrier functions and safe gradient flow. 

\subsection{Notation and Definitions} 
Given $f: \mathbb{R}^n \to \mathbb{R}^m$, let $\frac{\partial f}{\partial x} \in \mathbb{R}^{m \times n}$ be the Jacobian of $f(x)$. When $m = 1$, $\nabla_x f$ denotes the gradient. Consider a nonlinear program (NLP)
\begin{subequations}
\label{eq:nlp}
\begin{align}
\min_{x} \quad & c(x) \\
\textrm{s.t.} \quad & g(x) \leq 0 \\
& h(x) = 0
\end{align}
\end{subequations}
where $x \in \mathbb{R}^n$, $g: \mathbb{R}^n \to \mathbb{R}^l$, and $h: \mathbb{R}^n \to \mathbb{R}^k$.
Denote the feasible set as $\mathcal{C} = \{ x \in \mathbb{R}^n | g(x) \leq 0, h(x) = 0 \}$. 
Let $x^*$ denote a locally optimal point that satisfies the KKT conditions for \eqref{eq:nlp} and $X_{KKT}$ to be the set of all these points.



\subsection{Control barrier functions}
Recall the definition of a control barrier function from~\cite{ames_cbf_2019}. Given a control-affine system 
$
\dot{x} = \phi(x) + \gamma(x) u,
$
a function $b$ is a control barrier function (CBF) if
$$
\exists \alpha : \sup_{u} (\nabla_x b \cdot \phi(x) + \nabla_x b \cdot \gamma(x) \cdot u) \geq -\alpha(b(x)),
$$
where $\alpha$ is an extended class $\kappa$ function. If $b$ is a CBF and $\nabla_x b(x) \ne 0$ for $x \in \partial \mathcal{C}$ then any $u: \mathbb{R}^n \to \mathbb{R}^m$ satisfying $u(x) \in \{\mu: \dot{b}(x,\mu) + \alpha(b(x)) \geq 0\}$ renders $\mathcal{C}$ safe. 

\subsection{Safe gradient flow}
One approach for solving (\ref{eq:nlp}) that is guaranteed to return a feasible solution regardless of when it is terminated, is referred to as \emph{safe gradient flow} \cite{cortes_safeflow_2024}. This approach is shown to be asymptotically stable with equilibria in $X_\mathrm{KKT}$, if the constraints satisfies the so called Linear Independence Constraint Qualification Condition. 
Consider the control-affine system
\begin{equation}
\label{eq:sgf-dynamics}
\dot{x} = -\nabla_x c(x) - \frac{\partial g(x)}{\partial x}^\top \mu - \frac{\partial h(x)}{\partial x}^\top \nu,
\end{equation}
where $x \in \mathbb{R}^n$ is the state, and $\mu \in \mathbb{R}^l, \nu \in \mathbb{R}^k$ are inputs. Here the gradient of $c$ pushes $x$ in a direction that minimizes the cost function, while the additional inputs are used to ensure that the state remains within the safe set $\mathcal{C}$.

Applying CBF theory here, the admissible control set that ensures $\mathcal{C}$ is invariant is
\begin{align*}
\begin{split}
K(x) & =  \{(\mu,\nu) \hspace{0.5em}| \\ &
- \frac{\partial g}{\partial x} \frac{\partial g}{\partial x}^\top \mu - \frac{\partial g}{\partial x} \frac{\partial h}{\partial x}^\top \nu \leq \frac{\partial g}{\partial x} \nabla c(x) - \alpha(g(x)), \\ &
- \frac{\partial h}{\partial x} \frac{\partial g}{\partial x}^\top \mu - \frac{\partial h}{\partial x} \frac{\partial h}{\partial x}^\top \nu = \frac{\partial h}{\partial x} \nabla c(x) - \alpha(h(x))  \}.
\end{split}
\end{align*}
A feedback controller to find minimally altering values of $\mu, \nu$ is synthesized as the quadratic program (QP)
\begin{equation}
\label{eq:sgf-qp}
(\mu(x), \nu(x)) = \underset{\mu,\nu \in K(x)}{\mathrm{argmin}} \left \{ \left\| \frac{\partial g(x)}{\partial x}^\top \mu + \frac{\partial h(x)}{\partial x}^\top \nu \right\|^2 \right\}
\end{equation}
This feedback controller finds the minimum alteration to the gradient descent term required to make the set $\mathcal{C}$ invariant while driving the state to a locally optimal point.

\section{System Model} 
\label{sec:model}
\begin{figure}
    \begin{center}
\begin{circuitikz} [american voltages]
\draw
    (0,0) to node[tlground]{} (0,0)
    to[sV] (0,2)
    (0,2) node[circ,label=above:$\mathbf{V}_{\mathrm{dq}}$]{}
    to (0,2)--(0.5,2) 
    
    to [short, i=$\mathbf{I}_{\mathrm{dq}}$] (1.5,2)
    
    to[L, l_=$L$] (2.5,2)
    to[R, l_=$R$] (4,2)

    to [short, i=${P,Q}$] (5,2)

    to (5,2) -- (5.5,2) 
    (5.3,1.85) -- (5.3,2.15)
    (5.3,2) node[circ,label=above:$\mathbf{E}_{\mathrm{dq}}$]{}
    (5.6,2) -- (5.7,2)
    (5.8,2) -- (5.9,2)
    (6.0,2) -- (6.1,2);
\end{circuitikz}
\end{center}
    \caption{The dq reference frame model of an inverter connected to an infinite bus via an \emph{RL} branch, with the inverter modeled as a controllable voltage source. The reference frame is taken with respect to the inverter voltage angle.}
    \label{fig:circuit}
    \vspace{-5pt}
\end{figure}
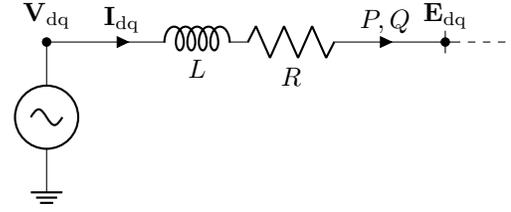

In this paper, we consider a simplified model of a three-phase inverter connected to an infinite bus via an \emph{RL} branch as shown in Fig.~\ref{fig:circuit}. 
The model assumes we have balanced three phase on the AC-side and can use a direct-quadrature (dq) reference frame with reference to the inverter voltage which rotates with frequency $\omega$~\cite{yazdani2010voltage}. This model is commonly used when studying the transient stability of inverters~\cite{Huang_Xin_Wang_Zhang_Wu_Hu_2019, joswig2024optimal, joswig2024safe}.
The system dynamics are
\begin{align}
    \label{eq:dynamics}
    f(x,u) = \begin{bmatrix}
        \dot{I_\mathrm{d}} \\
        \dot{I_\mathrm{q}} \\
        \dot{\delta} \\
    \end{bmatrix} = 
    \begin{bmatrix}
        -\frac{R}{L} I_{\mathrm{d}} + \omega  I_{\mathrm{q}} + \frac{\sqrt{2}}{L} (V - E \cos{(\delta)}) \\
        -\frac{R}{L} I_{\mathrm{q}} - \omega I_{\mathrm{q}} + \frac{\sqrt{2}}{L} ( - E \sin{(\delta)}) \\
        \omega_e - \omega
    \end{bmatrix}.
\end{align}
$\mathbf{I}_\mathrm{dq} = (I_\mathrm{d}, I_\mathrm{q})$ and $\mathbf{V}_{\mathrm{dq}}=(\sqrt{2} V,0)$ are the inverter current and voltage, respectively, in the inverter reference frame. The grid voltage rotates with frequency $\omega_e$ and $\delta$ is the angle difference between the grid voltage and inverter voltage. $\mathbf{E}_{\mathrm{dq}}=(\sqrt{2} E\cos(\delta),\sqrt{2} E\sin(\delta))$ defines the grid voltage in the inverter reference frame. 
We take the states to be $x = (I_\mathrm{d}, I_\mathrm{q}, \delta)$ and the control inputs to be $u = (V, \omega)$.
$P, Q$ are the active and reactive powers injected into the grid, computed as $ P = \frac{3}{2} \left( E \cos{(\delta)} I_\mathrm{d} + E \sin{(\delta)} I_\mathrm{q}  \right)$, $ Q = \frac{3}{2} \left( E \sin{(\delta)} I_\mathrm{d} - E \cos{(\delta)} I_\mathrm{q} \right)$.
To prevent the semiconductor switches of the inverter from becoming damaged, the magnitude of the current must be limited~\cite{luo2017advanced,bottrell2013comparison}:
\begin{equation}
    \label{eg:current-mag-lim}
    g(x) = \|\mathbf{I}_\mathrm{dq}\|_2^2 - I_\mathrm{max}^2 \leq 0.
\end{equation}

We want to control the inverter such that we inject close to some given desired active and reactive power values $P^*, Q^*$ without significantly deviating from the nominal voltage and frequency values $V_\mathrm{nom}, \omega_\mathrm{nom}$. We formulate our cost as
\begin{align} \label{eqn:obj}
\begin{split}
    c(x,u) = &\frac{m_p}{2}(P - P^*)^2 + \frac{m_q}{2\cdot \tau_v}(Q - Q^*)^2 \\ &+ \frac{1}{2\cdot \tau_v}(V - V_\mathrm{nom})^2 + \frac{1}{2}(\omega - \omega_\mathrm{nom})^2
\end{split}
\end{align}
where $m_p, m_q, \tau_v$ are the active power droop term, reactive power droop term, and voltage control speed term, respectively. 

Our goal is to define a controller that drives the system to optimal equilibrium points, which is equivalent to solving:
\begin{subequations}
\label{eq:inv-nlp}
\begin{align}
\min_{x,u} \quad & c(x,u) \\
\textrm{s.t.} \quad & g(x) \leq 0 \\
& f(x,u) = 0.
\end{align}
\end{subequations}
We note that we cannot directly apply the SGF approach here as our dynamical system~(\ref{eq:dynamics}) does not directly align with~(\ref{eq:sgf-dynamics}) and we may not always be able to fully control $f(x,u)$ such that it flows according to the gradient of the cost.

\section{Safe Gradient Flow for Dynamical Systems} 
\label{sec:stgf}
We design a controller using the \emph{safe gradient flow} method by applying it to a rolling horizon trajectory optimization problem. We discretize~\eqref{eq:inv-nlp} to obtain the following trajectory optimization problem:
\begin{align}
\begin{split}
\min_{x, u} \quad & c_f(x_{t+T}) + \sum_{\tau=t}^{t+T} c(x_\tau, u_\tau) \\
\textrm{s.t.} \quad & g(x_{\tau}) \leq 0 \hspace{0.5em} \forall \hspace{0.5em} \tau \in (t, t+T] \\
& h(x_\tau) = 0 \hspace{0.5em} \forall \hspace{0.5em} \tau \in (t, t+T] \\
& x_{\tau + 1} = f(x_{\tau},u_{\tau}) \hspace{0.5em} \forall \hspace{0.5em} \tau \in (t, t+T-1]
\end{split}
\end{align}
where $T$ is the horizon of the trajectory being optimized, and $x \in \mathbb{R}^{n \times T}, u \in \mathbb{R}^{m \times T}$ are the trajectories of the states and inputs, respectively. $c_f(x)$ is an altered cost function that only depends on the final state on the horizon $T$ by removing the terms with the inputs $V, \omega$. Next we denote
$$G(w) = \begin{bmatrix}
g(x_{t}) \\
\vdots \\
g(x_{t+T})
\end{bmatrix}
, H(w) =
\begin{bmatrix}
h(x_{t}) \\
\vdots \\
h(x_{t+T}) \\
x_{t + 1} - f(x_{t}, u_t) \\
\vdots \\
x_{t + T} - f(x_{t + T - 1}, u_{t + T - 1}) \\
\end{bmatrix},
$$
and $C(w) =  c_f(x_{t+T}) + \sum_{\tau=t}^{t+T} c(x_\tau, u_\tau)$, where $w =(x,u)$.
%
With this notation we can fit this into the form of~(\ref{eq:nlp}) as
\begin{subequations}
\label{eq:qp-stgf}
\begin{align}
\min_{w} \quad & C(w) \\
\textrm{s.t.} \quad & G(w) \leq 0 \\
& H(w) = 0
\end{align}
\end{subequations}
Applying the safe gradient flow method to this trajectory optimization problem steps the trajectory $(x,u)$ towards a locally optimal trajectory $(x^*,u^*)$. However, we want to implement a controller in real time and may only be able to updated the trajectory for a limited number of times per step. 

To solve this in real time and apply the control inputs to our system, we adopt a rolling horizon implementation inspired by model predictive control. At each time step, we formulate the above optimization problem using the current state as the initial condition, perform a limited number of updates, $K$, (i.e. gradient steps) toward minimizing the cost subject to constraints, apply the first control action from this solution, shift the horizon forward, and repeat the process. This process is detailed in Algorithm~\ref{alg:stgf}, which we call a \emph{safe trajectory gradient flow} (STGF) controller.
\begin{algorithm}
\caption{ Safe Trajectory Gradient Flow}
\label{alg:stgf}
\begin{algorithmic}[1]
\Require Initial state $x_0 \in \mathbb{R}^n$, Initial input trajectory $u \in \mathbb{R}^{(T-1) \times m}$, horizon length $T$, termination time $N$, optimization updates per step $K$, dynamics $f(x,u)$, constraints $(g(x,u), h(x,u))$, cost function $c(x,u)$, time step $\Delta t$, optimization step-size $\xi$, CBF parameter $\alpha$
\State Initialize trajectory: $x \in \mathbb{R}^{T \times n} \gets (x_0, f(x_0, u_0),\dots,f(x_{T-1}, u_{T-1}))$
\State Create $C, G, H$ from applying $c,g,h,f$ at each timestep in the horizon $T$ 
\While{$t \leq N$}
    \State Measure current timestep state $x_t$
    \State $x_0 \gets x_t$
    \State Propagate state trajectory with predicted dynamics: 
    \For{$\tau = 0$ to $T - 1$}
        $x_{\tau+1} \gets f(x_\tau, u_\tau)$
    \EndFor
    \For{$j = 0$ to $K$}
        \State $w \gets (x, u)$
        \State $\mu, \nu \gets$ Solve QP~(\ref{eq:sgf-qp}) with $w,C,G,H,\alpha$
        \State $\dot{w} \gets -D_w c(w) - \frac{\partial g(w)}{\partial{w}}^\top \mu - \frac{\partial h(w)}{\partial{w}}^\top \nu$
        \State $x,u \gets w + \xi \cdot \dot{w}$
    \EndFor
    \State Apply control input $u_0$ from the optimized trajectory
    \State Roll forward control inputs: 
    \For{$\tau = 0$ to $T-2$}
        $u_{\tau} \gets u_{\tau+1}$
    \EndFor
    \State $u_{T-1} \gets u_{T-2}$
    \State $t \gets t + 1$
\EndWhile
\end{algorithmic}
\end{algorithm}

This results in a feedback control policy that iteratively improves its trajectory while always maintaining feasibility (retaining the anytime property central to safe gradient flows), if the predicted state dynamics align with the true state dynamics
\footnote{Measuring exogenous values and estimating the system's true dynamics are important challenges with this approach, which we reserve for future work.} and a sufficiently small step size is used. 
Although the full optimal trajectory may not be reached at each step, the quality of the control inputs improves over time, and the feasible set remains invariant throughout.


\section{Simulation Results} 
\label{sec:results}
We test our STGF controller by apply the approach to the simplified inverter system with the per unitized parameters shown in Table~\ref{tab:parameters} and compare it to model predictive control and a modified droop controller.
\begin{table}
    \centering
    \caption{Simulation Parameters}
    \begin{tabular}{cc|cc}
    \hline
         Parameter & Value & Cost Parameter & Value\\ 
         \hline
         $R$& $0.0069~\text{pu}$ & $m_p$ & $2 \pi$ \\
         $L$ & $0.0196~\text{pu}$ & $m_q$ & $0.05$ \\
         $I_\mathrm{max}$ & $1~\text{pu}$ & $\tau_p$ & $0.1$ \\
         \cline{1-4}
         Base & Value & Opt. Parameter & Value \\
         \cline{1-4}
         $\omega_\mathrm{base}$ & $2\pi60$~\unit{\radian\per\second} & $\Delta t$ & $1~\unit{\milli\second}$ \\
         $V_\mathrm{base}$ & $120~\unit{\volt}$ & $\xi$ & $1\times 10^{-3}$ \\
         $S_\mathrm{base}$ & $1500~\unit{\volt\ampere}$ & $T$ & 10 \\
         $I_\mathrm{base}$ & $4.167~\unit{\ampere}$ & $K$ & 2 \\
         \hline
    \end{tabular}
    \label{tab:parameters}
\end{table}
To apply our STGF approach we discretize our dynamics (\ref{eq:dynamics}) using forward Euler with $\Delta t = 1~\unit{\milli\second}$. 
A control horizon of $T=10$ is selected and each time step only a single iteration of the trajectory optimization problem is performed with $K=2$ and $\alpha(z)=20 e^{10 z}$. The STGF QP (\ref{eq:sgf-qp}) is formulated with $\texttt{CVXPY}$~\cite{diamond2016cvxpy} and solved using \texttt{OSQP}~\cite{osqp}. The dynamics of the inverter system are simulated using the $\texttt{solve\_ivp}$ function from $\texttt{SciPy}$~\cite{2020SciPy-NMeth}. We compare our approach to model predictive control (MPC) implemented with $\texttt{do-mpc}$~\cite{Fiedler2023dompc, wachter2006ipopt, Andersson2019casadi}, and a droop controller modified with a current limiter and an additional synchronization term~\cite{joswig2024optimal, Baeckeland2023}. We assume the controllers have perfect knowledge of the grid voltage and frequency in each timestep.
The code used to produce these results will be made available at \url{https://github.com/TragerJoswig-Jones/Safe-Trajectory-Gradient-Flow}.

\subsection{Trajectory Performance \& Steady-State Optimality}
\begin{figure}
    \centering
    \includegraphics[width=1.0\columnwidth, trim={0 0 0 0},clip]{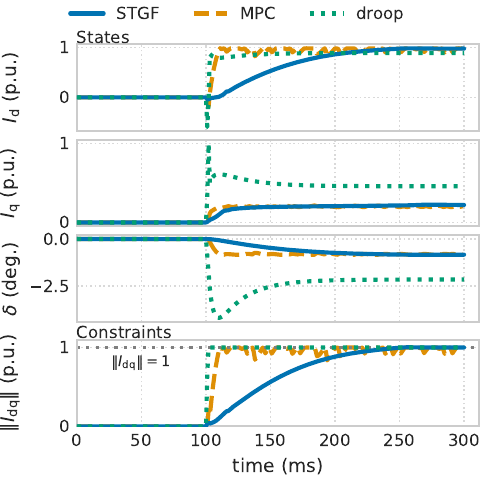}
    \vspace{-15pt}
    \caption{The states and current magnitude from the inverter system controlled with different control methods. While the MPC and droop controllers act are more aggressive, the STGF control drives the states smoothly to their new equilibria.}
    \label{fig:xu_traj_t}
    \vspace{-10pt}
\end{figure}

\begin{figure}
    \centering
    \includegraphics[width=\columnwidth, trim={0 0 0 0},clip]{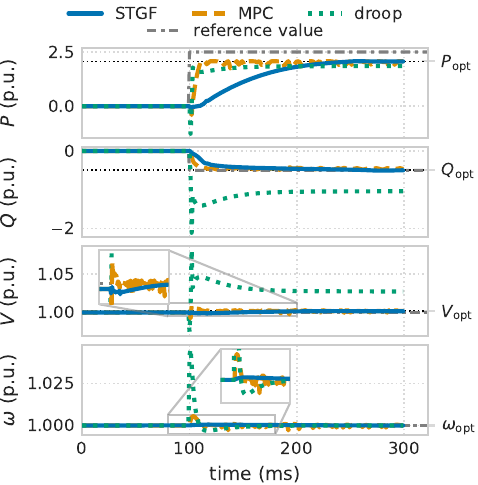}
    \vspace{-15pt}
    \caption{The outputs of the system included in the cost function for each control method. With STGF or MPC control the outputs converge to optimal values 'near' the given reference values. The droop control converges to values away from the optimal values.}
    \label{fig:y_traj}
    \vspace{-5pt}
\end{figure}

We test the system by simulating a step in the active and reactive power reference. The simulation lasts $N = 300$ time steps (i.e. $0.3~\unit{\second}$) with initial values $x_0=\mathbf{0}, u_0=\mathbf{0}$, and reference values $P^* = 0~\text{p.u.}, Q^*=0~\text{p.u.}, V_\mathrm{nom}=1~\text{p.u.}, \omega_\mathrm{nom}=2\pi \cdot 60~\unit{\radian\per\second}$. At $T=100$, we step the active and reactive power references to infeasible values $P^* = 2.5~\text{p.u.}, Q^*=-0.5~\text{p.u.}$. Fig.~\ref{fig:xu_traj_t} shows the trajectory of $\mathbf{I}_\mathrm{dq}$ with the STGF controller approaching the desired reference values while remaining within the current magnitude constraint. 
In Fig.~\ref{fig:y_traj}, the output cost parameters are seen to converge to optimal values calculated by solving (\ref{eq:inv-nlp}) directly using a NLP solver. 
The MPC controller also approaches the optimal values and does so faster than the STGF controller, but struggles to maintain consistent control inputs when the nonlinear current limit constraint is active. The droop controller also acts faster, but does not settle to the optimal outputs. 

Ideally, one would be able to select a horizon within which the system would be able to settle to a new equilibrium within. However, there is a trade-off that must be made in terms of computation time and numerical errors, and the length of the trajectory $T$, number of optimizing steps $K$, step size $\xi$, and the selected discretization time step $\Delta t$. These are important tuning-parameters to investigate when applying this approach. These simulations show that with a long enough horizon the future inputs can be continuously optimized as the systems state evolves even with only a single optimization step each time step of the simulation. When selecting $T$ it is important to ensure that $T > r$, were $r$ is the relative degree of the system dynamics which is the number of times the output values used in $c(x,u)$ must be differentiated before the input explicitly appears. 
For MPC one must select similar parameters to STGF relating to the trajectory optimization problem formulation. With the augmented droop controller there are multiple parameters in the synchronization term that must be tuned to ensure the droop controller remains stable, which can be challenging to guarantee for all cases.

\subsection{Controller Computational Burden \& Solve-time}

%

%

\begin{figure}
    \centering
    \includegraphics[width=\linewidth, , trim={0 0 0 0},clip]{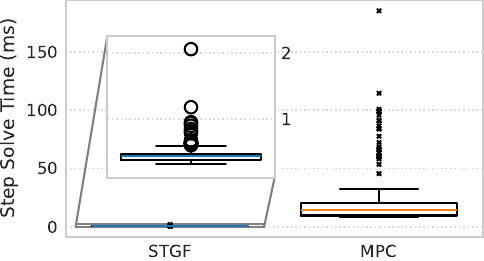}
    \vspace{-10pt}
    \caption{Single-step run-times for solving controller optimization problems. An inset enlarged view of STGF run-times shows that they are significantly shorter than most MPC run-times. }
    \label{fig:runtime}
    \vspace{-10pt}
\end{figure}

Although QPs are considered to be simple optimization problems, using iterative solvers in real-time on inverters themselves is not trivial, because of microprocessor limitations and sampling/input delays~\cite{Singletary_Chen_Ames_2020, Buso_Mattavelli_2015}. 
For these test we run the controllers and simulations using an Intel(R) Xeon(R) 2.20GHz CPU, which we note has significantly more processor power than a standard microprocessor. Fig.~\ref{fig:runtime} shows the optimization solve times for each step of the simulation for the STGF and MPC controllers. The average STGF controller single step runtime is roughly 40 times faster than the MPC controller, demonstrating its potential to generate faster trajectories updates compared to fully solving the nonlinear trajectory optimization problem each time step. Further, the maximum solve time of the STGF QP is orders of magnitude smaller than that of MPC. This difference in solve time is most pronounced when the current magnitude limit is active. The average solve time of STGF was still greater than the simulation step sized used in these tests, indicating that without further computational speed up these approaches could not be used directly to control an inverter system with this setup. In comparison the explicit droop controller implementation is significantly less computationally intensive. 
%

\section{Conclusion} \label{sec:conclusion}
%
%

In this paper we proposed a feedback controller for dynamical systems with state constraints that optimizes a nonlinear program. Our approach applies the safe gradient flow method to a grid-interfacing inverter system by formulating it as a rolling horizon problem and demonstrates that the approach can drive the system towards an optimal operating point. 

Our future work include finding conditions to guarantee the convergence of the rolling horizon trajectory problem, and  how to handle the impact of disturbances on the ability of the approach to maintain safety. In addition, reducing the computational complexity to make it easier to implement on microcontrollers is an important topic.

\bibliographystyle{IEEEtran}
\bibliography{Reference}
\end{document}